\documentstyle{article}
\begin{document}
\input epsf
\rightline{UM-P-95/115}
\rightline{RCHEP 95/27}

\vskip 1.2cm
\begin{center}
\begin{large}
{\bf Electric charge quantisation from gauge invariance of
a Lagrangian: A catalogue of baryon number violating scalar interactions.}
\vskip 1.5cm
\end{large}
J. P. Bowes, R. Foot
and R. R. Volkas
\end{center}
\vskip 1cm
\noindent
Research Centre for High Energy Physics\\
School of Physics\\
University of Melbourne\\
Parkville, 3052\\
Australia.

\vskip 2cm
\begin{center}
{\bf Abstract}
\end{center}
\vskip 1cm
In gauge theories like the standard model, the electric charges of the
fermions can be heavily constrained from the classical structure of
the theory and from the cancellation of anomalies.
There is however mounting evidence suggesting that these anomaly
constraints are not as well motivated as the classical constraints. In
light of this
 we discuss possible modifications of the minimal standard
model which will give us complete electric charge quantisation from  
classical
constraints alone.
Because these modifications to the Standard Model involve the consideration of 
baryon number violating scalar interactions, we present a 
complete catalogue of the simplest ways to modify the Standard Model so as 
to introduce explicit baryon number violation. This has implications for 
proton decay searches and baryogenesis.

\newpage

\begin{center}
{\bf 1. Introduction and Motivation}
\end{center}

Investigation of explicit baryon number violation in simple extensions 
of the Standard Model (SM) is interesting for a number of different reasons, 
including: a) the requirement of baryon number violation to explain 
baryogenesis; and  b) the continuing interest in terrestrial searches for 
 baryon number violating processes. 

The aim of this work is to provide a complete catalogue of the simplest 
ways to explicitly violate baryon number through extensions of the SM. 
A theoretical motivation for doing this arises also from the work done by 
one of us \cite{main} on the possibility of obtaining complete electric 
charge quantisation 
from classical constraints. 

The quantisation of the electric charges of the known fermions
is a well established experimental phenomenon. An approach to
a theoretical understanding of this phenomenon
has emerged in recent years based on the SM [1].
The SM is a gauge theory with gauge group
\begin{equation}
SU(3)_c \otimes SU(2)_L \otimes U(1)_Y,
\end{equation}
which is assumed to be spontaneously broken by the vacuum
expectation value (VEV) of a scalar doublet $\phi \sim (1, 2, 1)$.
The $U(1)_Y$ charge of $\phi$ can be normalised to $1$ without loss of generality due to a scaling symmetry, $g \rightarrow \eta g, Y \rightarrow Y/\eta$, where $g$ is the $U(1)_Y$ coupling constant, and $Y$ is the generator of the
$U(1)_Y$ gauge group. The gauge symmetry of the Lagrangian can be used to
choose the standard form for the vacuum:
\begin{equation}
\langle \phi \rangle = \left(\begin{array}{c}
0\\
u
\end{array}\right).
\end{equation}
The VEV of $\phi$ breaks $SU(2)_L \otimes U(1)_Y$ leaving
an unbroken $U(1)$ symmetry, $U(1)_Q$, which is identified with
electromagnetism. Its generator $Q$
is the linear combination which annihilates the VEV of Eq.(2):
\begin{equation}
Q = I_3 + Y/2.
\end{equation} 
The normalisation of $Q$ is not physically measurable, and we have adopted the convention of normalising it so that the charged $W$ bosons will have charge 1.
The above reasoning shows that the electric charge quantisation problem
would be solved if a way could be found to deduce the Y-charges of the
fermions.

There are two quite distinct ways in which the standard model
constrains the electric charges of the fermions.
First, there are a set of constraints which follow from
the definition of the theory at the classical level:
the requirement
that the Lagrangian be gauge invariant. 
Second, there are other
constraints which are assumed to 
follow from the consistency of the theory at the
quantum level: the anomaly cancellation conditions.
The outcome of this is that charge quantisation follows
provided that there is only one anomaly-free U(1) symmetry of the
Lagrangian outside of those contained in $SU(3)_c\otimes SU(2)_L$. If it turns out that the
generator of this U(1) symmetry is 
precisely standard weak-hypercharge $Y$, then not only is
charge quantised but it is quantised correctly. 

For instance, consider the minimal SM. In addition
to standard $Y$, any one of $L_e - L_{\mu}$, $L_e - L_{\tau}$ and
$L_{\mu} - L_{\tau}$ generates an anomaly-free U(1) symmetry of the
Lagrangian. Therefore the minimal SM poses a charge quantisation
problem because the actual weak-hypercharge of the theory can
be chosen to be $\cos\Theta Y_{standard} + \sin\Theta (L_i - L_j)$
where $\Theta$ is an arbitrary parameter and $i,j = e, \mu, \tau (i\neq
j)$. See Ref.\cite{refxx} for more detailed reviews.

The above analysis assumes that the cancellation of gauge anomalies
is a rigorous requirement for a consistent gauge theory. 
There are however several arguments which throw doubt on the validity
 of this requirement. For example 
there may be a set of as yet undetected mirror fermions which remove the
anomaly cancellation requirement. There are also interesting arguments 
given by Kieu
\cite{kieu} in a series of papers to the effect that a properly analysed
``anomalous'' gauge theory is not anomalous at all.
(For other interesting work on the question of the
consistency or otherwise of anomalous gauge theory see Ref.\cite{jackiw}.)
If gauge
anomaly cancellation as routinely enforced is unnecessary, then
there is no motivation to use these constraints in deriving
electric charge quantisation. Clearly one is then left with
the following result: {\it Electric charge quantisation will be a
necessary outcome of the construction of a theory (i.e. a Lagrangian)
provided that it displays only one unembedded U(1) invariance. If the generator
of this single U(1) symmetry is standard weak-hypercharge, then not only
is charge quantised but it is quantised correctly.}

The three-generation minimal SM has five U(1) invariances [aside 
from U(1) subgroups of SU(3)$\otimes$SU(2)]. In addition to
standard weak-hypercharge, there is baryon number $B$ and the three
family lepton-numbers $L_e$, $L_{\mu}$ and $L_{\tau}$. If gauge anomaly
cancellation is not enforced, then the generator of the gauged U(1) in
the minimal SM can be any linear combination of $Y$, $B$ and the
$L_i$. This leads to a four-parameter charge quantisation problem.

These simple observations provide strong motivation to construct
extensions of the minimal SM that explicitly break $B$ and the $L_i$
(but of course leave $Y$ exact). All such models would explain charge
quantisation in the sense that they simply could not be constructed 
unless charge was quantised (i.e. some terms in the Lagrangian would
have to be absent in order to reinstate $B$ or any of the $L_i$ as a
conserved charge). The purpose of this paper is to construct the
simplest extensions of the minimal SM that explicitly break $B$ and each
of the $L_i$. Further, we will examine the most stringent
phenomenological constraints on these models and thus determine those
that are least constrained and hence of most experimental interest. This
type of analysis was first performed in detail in Ref.\cite{main}.
We will extend the analysis of Ref.\cite{main} and correct an important 
technical error. This is also a motivation for the present work.

The four parameter charge quantisation problem of the minimal SM
corresponds to there being
four classically undetermined electric charges, 
which can be taken to be the three neutrino charges 
and the down quark charge \cite{main}. From experimental data we 
know that three of these four charges are strongly 
constrained \cite{limit}, with only $Q(\nu_{\tau})$ being weakly constrained 
\cite{main,fandl}. In the following work
we seek to remove this four parameter uncertainty by means of simple
extensions of the minimal standard model which explicitly break
$U(1)_B$ and each of the $U(1)_{L_i}$.

The simplest and most phenomenologically interesting way to explicitly
break the $U(1)_{L_i}$ is to introduce non-zero neutrino masses. This is
most easily done by introducing right handed neutrinos into the model.
If we choose that our right and left handed neutrinos are related
through Dirac mass terms,
${\cal L}=\lambda \bar{\nu}_L \nu_R+H.c$.,
and if we assume that nontrivial mixing effects occur 
as in the quark sector, then we obtain the
constraint $Q(\nu_e)=Q(\nu_{\mu})=Q(\nu_{\tau})$. This leaves just two
undetermined electric charges, which can be taken to be $Q(\nu_e)$ and
$Q(d)$, corresponding to the as yet unbroken global symmetries $U(1)_L$
 and $U(1)_B$ where $L=L_e+L_{\mu}+L_{\tau}$ is total lepton number. If we then add a Majorana mass term, ${\cal L}=\lambda \bar{\nu}_R
(\nu_R)^c$, for one or more of the right
handed neutrinos we obtain the additional constraint
$Q(\nu_e)=0$ \cite{bandm}. Or put another way, the Majorana mass terms explicitly
break $U(1)_L$. This leaves just one undetermined electric charge, which
can be taken to be the electric charge of the down quark, $Q(d)$.
Our four parameter uncertainty has therefore been reduced to a one parameter
uncertainty by this simple extension of the lepton sector. 

Our remaining global symmetries are the hypercharge $U(1)_Y$ and the baryon number
$U(1)_B$. Hence, assuming that anomaly cancellation is unnecessary, any combination of $Y$ and $B$
can be the $U(1)$ symmetry which is gauged. To obtain complete electric charge
quantisation we require that this unwanted baryon number symmetry somehow be broken without affecting the $U(1)_Y$ hypercharge symmetry. This double
requirement rules out the introduction of baryon number violating Majorana quark mass terms, because  
unlike their lepton counterparts such terms will result in the violation of
standard hypercharge and colour.

To achieve charge quantisation in the simplest way, using only
the standard model gauge symmetry, we therefore require the addition of
a new scalar which incorporates the dual requirements of baryon number 
violation and hypercharge conservation \cite{main}. The violation of baryon
number requires that this new scalar interact with quarks, and
 assuming the usual dimension four (Yukawa-type) couplings
there is a finite list of possible quantum numbers for this scalar.
Since the scalar couples to a fermion bilinear, it follows from 
gauge invariance that the quantum numbers of the scalar are those of the
fermion bilinears. For example a scalar $\sigma_1$ coupling via the
interaction term ${\cal L}=\lambda\sigma_1^{\dagger} \bar{Q}_L(f_L)^c$
implies that $\sigma_1$ transforms as $\bar{Q}_L(f_L)^c$. Following such a procedure all possible scalars in terms of fermion
bilinears can be found (see Ref.\cite{main}). These scalars 
together with their $SU(3)_C\otimes SU(2)_L\otimes U(1)_Y$ 
representations are listed below:
\begin{eqnarray}
\begin{array}{lclcl}
\sigma_{1.1}& \sim& \bar{Q}_L(f_L)^c\sim\bar{u}_R(e_R)^c\sim\bar{d}_R(\nu_R)^c&
 \sim& (\bar{3},1,-y_d)(-1/3)\\
\sigma_{1.2}& \sim& \bar{Q}_L(f_L)^c& \sim& (\bar{3},3,-y_d)(-1/3)\\ 
\sigma_2& \sim& \bar{Q}_Le_R\sim\bar{u}_Rf_L& \sim& (\bar{3},2,-3-y_d)(-1/3)\\
\sigma_{3.1}& \sim& \bar{Q}_L(Q_L)^c\sim\bar{u}_R(d_R)^c &
 \sim& (3,1,-2-2y_d)(-2/3)\\ 
\sigma_{3.2}& \sim& \bar{Q}_L(Q_L)^c& \sim& (3,3,-2-2y_d)(-2/3)\\
\sigma_{3.3}& \sim& \bar{Q}_L(Q_L)^c\sim\bar{u}_R(d_R)^c &
 \sim& (\bar{6},1,-2-2y_d)(-2/3)\\
\sigma_{3.4}& \sim& \bar{Q}_L(Q_L)^c& \sim& (\bar{6},3,-2-2y_d)(-2/3)\\
\sigma_4& \sim& \bar{u}_R(\nu_R)^c& \sim& (\bar{3},1,-2-y_d)(-1/3)\\
\sigma_5& \sim& \bar{d}_Rf_L\sim\bar{Q}_L\nu_R& \sim& (\bar{3},2,-1-y_d)(-1/3)\\
\sigma_{6.1}& \sim& \bar{u}_R(u_R)^c& \sim& (3,1,-4-2y_d)(-2/3)\\
\sigma_{6.2}& \sim& \bar{u}_R(u_R)^c& \sim& (\bar{6},1,-4-2y_d)(-2/3)\\
\sigma_{7.1}& \sim& \bar{d}_R(d_R)^c& \sim& (3,1,-2y_d)(-2/3)\\
\sigma_{7.2}& \sim& \bar{d}_R(d_R)^c& \sim& (\bar{6},1,-2y_d)(-2/3)\\
\sigma_{8}& \sim& \bar{d}_R(e_R)^c& \sim& (\bar{3},1,2-y_d)(-1/3).
\end{array}
\end{eqnarray}
Note that we have included the baryon number of the fermion bilinear
with which each scalar interacts as the last entry in each line above and 
we have used the following notation for the standard model fermions
and right handed neutrinos:
\begin{eqnarray}
&f_L&\sim(1,2,-1),\;\;\;\;\;e_R\sim(1,1,-2),\;\;\;\;\;\nu_R\sim(1,1,0),
\nonumber\\
&Q_L&\sim(3,2,1+y_d),\;\;\;\;u_R\sim(3,1,2+y_d),\;\;\;\;
d_R\sim(3,1,y_d).
\end{eqnarray}
It should also be pointed that the fermion interactions, $\bar{Q}_L(Q_L)^c$, 
$\bar{Q}_L(Q_L)^c$, $\bar{u}_R(u_R)^c$, and $\bar{d}_R(d_R)^c$
 associated with the 
$\sigma_{3.2}$, $\sigma_{3.3}$, $\sigma_{6.1}$, and $\sigma_{7.1}$ scalars
are flavour antisymmetric. 

Due to the fact that these proposed scalar particles may carry
baryon number, the above interactions by
themselves will not violate baryon number. 
Instead we can break baryon number by either 
proposing the existence of more than one quark-lepton 
interaction, or alternatively 
by proposing the existence of two or more scalar multiplets together with 
their associated interactions. 

\begin{center}
{\bf 2. One scalar extensions}
\end{center}

In the interests of simplicity Ref.\cite{main} considered the
case where just one of these scalar particles existed, with $U(1)_B$
being broken explicitly in the Higgs potential. As a result it was found
that because all of the scalars are either in the {\bf 3} or {\bf 6}
representation of $SU(3)_c$, the only renormalisable terms which break
baryon number and conserve $SU(3)_c$ and hypercharge are $\sigma^3\phi$ or $\sigma^3\phi^{\dagger}$. 
Since the Higgs doublet $\phi$ has hypercharge 1 (in our
normalisation) these scalar potentials require that our scalar particle 
$\sigma$
has either a hypercharge of $-1/3$ or $1/3$ respectively. Out of all the
possibilities listed in Eq.(4) only $\sigma_5$ satisfies either of these 
constraints for the observed value of $y_d=-2/3$.
It was thus concluded in Ref.\cite{main}
that under the assumption of one exotic scalar and one set of quark
lepton interactions, that electric charge can
be quantised classically.

Upon closer examination it is however found that the scalar potential
term $\sigma_5^3\phi$ is in fact zero after
antisymmetrisation over the $SU(3)_c$ group (this is the error in Ref.\cite{main} alluded to earlier). We must therefore broaden our search for baryon number violating extensions to the standard model which give the desired charge quantisation.

We are primarily interested in simple extensions to the model. Thus we
will initially continue to search for extensions which require 
the introduction of just one scalar particle. However we know from 
the unsuccessful attempts made in
Ref.\cite{main} that the consideration of just one of the interactions
shown in Eq.(4) will not provide the required charge quantisation. In our
quest for charge quantisation we must therefore take the next step and
consider pairs of interactions in Eq.(4) which can couple to the same
scalar in a baryon number violating manner. 
Two different quark-lepton interactions can only couple to the
one scalar if the group properties of this scalar are compatible with
both interactions. If this compatibility is subject to the strict
condition that $y_d=-2/3$, then not only do we have two sets of
interactions associated with the one scalar, but we also obtain the
desired charge quantisation; i.e. the two sets of interactions provide
us with the baryon number violation required for charge quantisation.
An example of such a pair is $\sigma_{1.2}$ and 
$\sigma_{3.2}^{\dagger}$, which are
equal subject to the constraint $y_d=-2/3$ as desired. Thus $\sigma_{1.2}$ and 
$\sigma_{3.2}$
 can be conjugate representations of the same particle, call it 
$\sigma$, with the following lepton and quark interactions,
\begin{equation}
{\cal L}=\lambda_1 \overline{(f_L)^c}\sigma Q_L+\lambda_2 \bar{Q}_L
\sigma (Q_L)^c+H.c.
\end{equation}
Because $\overline{(f_L)^c}\sigma Q_L$ and $\bar{Q}_L
\sigma (Q_L)^c$ have different baryon numbers, $U(1)_B$ is explicitly
broken.

All together there are three possible conjugate pairs, which we list
below:
\begin{eqnarray}
\sigma_{1.1}&=&\sigma_{3.1}^c\sim(\bar{3},1,2/3),\nonumber\\
\sigma_{1.2}&=&\sigma_{3.2}^c\sim(\bar{3},3,2/3),\nonumber\\
\sigma_4&=&\sigma_{7.1}^c\sim(\bar{3},1,-4/3).
\end{eqnarray}

For each of the above conjugate pairs there will be a set of quark
lepton interactions obtainable from Eq.(4) [for example
the second pair has the interactions shown in Eq.(6)]. From these quark lepton 
interactions it is easy to see that 
baryon number violation of magnitude $\Delta B=1$ will occur, and as such these quark lepton interactions will give rise to nucleon decay [see Fig.1]. 
The simplest Feynman diagrams leading to nucleon decay
processes involve the production of one meson and
one antilepton, Fig.1,
 with decay width (evaluated from
dimensional considerations only) of the form
\begin{equation}
\Gamma
={\cal O}\left(\frac{\lambda^4 M_N^5}{M_{\sigma}^4}\right).
\end{equation}
$M_N$ represents the nucleon mass, $M_{\sigma}$ the scalar
particle mass, and the dimensionless $\lambda^4\equiv\lambda_1^2
\lambda_2^2$ factor represents the
contribution made by the Yukawa coupling constants see [Eq.(6)]. We can obtain a limit on the scalar mass by comparing the
above decay width with the experimentally known lower limit on
the proton lifetime  $\approx 10^{32}$years \cite{protondecay}. This
gives a lower limit on $M_{\sigma}$ of
\begin{equation}
M_{\sigma}>\lambda\times 10^{16}{GeV}. 
\end{equation}
This lower limit however does not hold true for the $\sigma_4-\sigma_{7.1}^c$ conjugate pair
containing the left handed antineutrino producing $\sigma_4$ scalar. The production of left handed 
antineutrinos will result in there being an extra suppression factor in Eq.(8), which will 
consequently give rise to a less strict lower limit on the scalar mass; this special case will be considered later.

\begin{center}
{\bf 3. Two scalar extensions}
\end{center}

The above lower limit on $M_{\sigma}$ (which was obtained by proposing a one scalar particle extension to the
standard model) is very large, and as such is phenomenologically
uninteresting. In light of this it is desirable to consider the existence of two
new scalar particles, in the hope of obtaining a 
phenomenologically more interesting model.
Although this extension will complicate our model by requiring
the introduction of scalar potential terms, there will still only be 
two sets of fermion-scalar interactions as in
Eq.(6). 

For example a two scalar model can be used to resolve the antisymmetrisation problem alluded to
earlier; i.e. the zero value of the scalar potential term
$\sigma_5^3\phi$. This is achieved by proposing the existence of two 
species of scalar with the same group transformations as $\sigma_5$, 
which we will call $\sigma_a$ and $\sigma_b$. These particles will 
couple to leptons and quarks through the Lagrangian, 
\begin{equation} 
{\cal L}=\lambda_1^a\bar{f}_L\sigma_a d_R+\lambda_1^b\bar{f}_L\sigma _b d_R+\lambda_2^a\bar{Q}_L\sigma_a^c\nu_R+\lambda_2^b\bar{Q}_L
\sigma_b^c\nu_R+H.c.,
\end{equation}
and will combine in the nonzero hypercharge constraining scalar potential term,
\begin{equation}
\Delta
V(\phi,\sigma_a,\sigma_{b})=\lambda\sigma_{a}\sigma_{b}^2\phi+
H.c.
\end{equation}

The $\sigma_a$ and $\sigma_b$ scalar pair is just one of many 
combinations of scalar particles in Eq.(4) which violate baryon number whilst
giving the correct hypercharge assignment to the down quark $y_d=-2/3$. 
However unlike the 
$\sigma_a$ and $\sigma_b$ pair, these other 
combinations will in general involve scalars with different group 
transformation properties.
By considering every possible scalar combination in Eq.(4),
two lists of possible charge quantising scalar potentials can be 
compiled, corresponding to 
$\Delta B=1$ baryon number violating processes and $\Delta
B=2$ baryon number violating processes respectively. The $\Delta B=1$ list
is shown below: 
\begin{eqnarray}
\sigma_1,\sigma_2&\rightarrow&\sigma_{1.2}\sigma_{1.2}\sigma_2\phi\nonumber\\
\sigma_1,\sigma_3&\rightarrow&\sigma_{1.1}\sigma_{3.1}+\sigma_{1.1}\sigma_{1.1}^c\sigma_{1.1}\sigma_{3.1}+\sigma_{1.1}\sigma_{3.1}\sigma_{3.1}^c\sigma_{3.1}
+\sigma_{1.1}\sigma_{3.1}\phi^{\dagger}\phi\nonumber\\
&\rightarrow&\sigma_{1.2}\sigma_{3.2}+\sigma_{1.2}\sigma_{1.2}^c\sigma_{1.2}
\sigma_{3.2}+\sigma_{1.2}\sigma_{3.2}\sigma_{3.2}^c\sigma_{3.2}
+\sigma_{1.2}\sigma_{3.2}\phi^{\dagger}\phi\nonumber\\
\sigma_1,\sigma_5&\rightarrow&\sigma_{1.1}\sigma_5\sigma_5 \nonumber\\
&\rightarrow&\sigma_{1.2}\sigma_{1.2}\sigma_5\phi^c\nonumber \\
\sigma_1,\sigma_6&\rightarrow&\sigma_{1.2}\sigma_{6.1}\phi\phi\nonumber\\
\sigma_1,\sigma_7&\rightarrow&\sigma_{1.2}\sigma_{7.1}\phi^c\phi^c\nonumber\\
\sigma_2,\sigma_3&\rightarrow&\sigma_2^c\sigma_{3.2}\sigma_{3.2}\phi^c\nonumber\\
\sigma_2,\sigma_7&\rightarrow&\sigma_2\sigma_{7.1}\phi\nonumber\\
\sigma_3,\sigma_4&\rightarrow&\sigma_{3.2}\sigma_4\phi\phi \nonumber\\
\sigma_3,\sigma_5&\rightarrow&\sigma_{3.1}\sigma_5\phi\nonumber\\
&\rightarrow&\sigma_{3.2}\sigma_{3.2}\sigma_5^c\phi\nonumber \\
\sigma_3,\sigma_{8}&\rightarrow&\sigma_{3.2}\sigma_{8}\phi^c\phi^c\nonumber\\
\sigma_4,\sigma_7&\rightarrow&\sigma_4\sigma_{7.1}+
\sigma_4\sigma_4^c\sigma_4\sigma_{7.1}+
\sigma_{7.1}\sigma_{7.1}^c\sigma_{7.1}\sigma_4
+\sigma_4\sigma_{7.1}\phi^{\dagger}\phi\nonumber\\
\sigma_5,\rho&\rightarrow&\sigma_5\sigma_5\sigma_5\rho\nonumber\\
\sigma_5^a,\sigma_5^b&\rightarrow
&\sigma_5^a\sigma_5^b\sigma_5^b\phi\nonumber\\
\sigma_5,\sigma_7&\rightarrow&\sigma_5\sigma_{7.1}\phi^c\nonumber\\
\sigma_6,\sigma_{8}&\rightarrow&\sigma_{6.1}\sigma_{8}+\sigma_{6.1}
\sigma_{6.1}^c\sigma_{6.1}\sigma_8+\sigma_8\sigma_8^c\sigma_8\sigma_{6.1}+
\sigma_{6.1}\sigma_8\phi^{\dagger}\phi
\end{eqnarray}
where $\phi$ represents the SM Higgs scalar $\phi\sim(1,2,1)$, and $\rho$
represents a new Higgs like scalar $\rho\sim(8,2,1)$.   
The scalar $\rho$ is a new non-standard model particle (like $\sigma$) 
which differs from the Higgs scalar in that it carries colour charge;
in fact it transforms as an {\bf 8} under $SU(3)_c$. This means that
unlike the Higgs particle $\phi$, $\rho$ will not take part in symmetry 
breaking and will not form a VEV. Apart from this colour structure, it
has similar Yukawa couplings to $\phi$.

The above scalar potential terms can be placed into groups consisting of
quadratic, cubic and quartic terms, with each
 subgroup giving rise
 to its own characteristic expression for the proton decay width.
By comparing these decay widths with the known lower limits on the nucleon
lifetimes, each subgroup will give its own particular constraint on the the 
masses of the scalar particles involved. We will again be using dimensional
arguments.

Note that in the following analysis we will initially be ignoring 
terms involving the
$\sigma_4$ scalar, due to the complications associated with the
production of the right handed neutrino and left handed antineutrino. 

Our analysis begins with the quadratic potential terms, $\mu^2\sigma_i
\sigma_j$ [see Eq.(12)], where $\mu^2$ is the coupling constant with dimensions of mass squared. These two particle interactions can be considered as constituting the off diagonal elements of the exotic scalar mass matrix. 
The simplest nucleon decay processes which can be obtained from
these bilinears involve the decay of the nucleon into a meson and an 
antilepton. For example the proton decay diagram resulting from the 
$\sigma_{1.1}\sigma_{3.1}$ bilinear
is shown in Fig.2. From dimensional arguments these bilinears give
rise to decay widths of the form:
\begin{equation}
\Gamma\simeq{\cal O}\left(\frac{\lambda^4\mu^4 M_N^5}{M_{\sigma_i}^4 M_{\sigma_j}^4}\right).
\end{equation}
In this case the dimensionless constant
$\lambda^4\equiv \lambda_i^2\lambda_j^2$, where $\lambda_i$ and
$\lambda_j$ 
represent the Yukawa couplings associated with $\sigma_i$ and $\sigma_j$.
If we compare this decay width with experimental lower limits on the proton life time, i.e. 
$\approx 10^{32}$years, we obtain a lower limit on $M_{\sigma}$ of:
\begin{equation}
M_{\sigma}>(\lambda\mu/M_{\sigma})\times 10^{16}GeV,
\end{equation}
where we have expressed
the coupling constant $\mu$ in terms of the scalar mass $M_{\sigma}$.
  
The simplest nucleon decay processes resulting from the Higgs doublet 
containing quadratic terms,
 $b\langle\phi\rangle\sigma_i\sigma_j$, involve the decay of a nucleon into a 
lepton and a meson [see Fig.3 for example]. 
The decay widths for these processes take the following form:
\begin{equation}
\Gamma\simeq {\cal O}\left(\frac{\lambda^4 b^2 \langle\phi\rangle^2 M_N^5}
{M_{\sigma_i}^4 M_{\sigma_j}^4}\right).
\end{equation}
In this case the coupling constant $b$ has units of mass and the 
Yukawa coupling constants have again been taken into consideration via the 
$\lambda^4\equiv \lambda_i^2\lambda_j^2$ factor. The above decay processes are
experimentally constrained by a lifetime lower limit of $\;\approx 10^{31}$yrs 
\cite{protondecay}, 
which gives rise to a constraint on $M_{\sigma}$ of
\begin{equation} 
M_{\sigma}>(\lambda^2 b/M_{\sigma})^{1/3}\times 10^{11}GeV.
\end{equation}

The simplest nucleon decay process resulting from the quadratic terms 
with two Higgs scalars, 
i.e. $\lambda\langle\phi\rangle\langle\phi\rangle\sigma_i\sigma_j$, 
involve the creation of a meson and an 
antilepton product. The decay widths for these processes 
take the following form:
\begin{equation}
\Gamma\simeq {\cal O}\left(\frac{\lambda^6\langle\phi\rangle^4 M_N^5}{M_{\sigma_i}^4 M_{\sigma_j}^4}\right),
\end{equation}
where the $\lambda^6$ factor represents the combined contribution made by the, 
$\lambda^4\equiv \lambda_i^2\lambda_j^2$ Yukawa coupling and the $\lambda^2$ exotic scalar
 coupling. These nucleon decay processes are best constrained by the
 $\tau_n> 10^{32}$ yr bound neutron decay limit 
\cite{protondecay}, giving rise to 
 the constraint
\begin{eqnarray}
M_{\sigma}>\lambda^{3/4}\times 10^9{GeV}. 
\end{eqnarray}
The above analysis is not however valid for the 
$\langle\phi\rangle\langle\phi\rangle\sigma_{1.2}\sigma_{6.1}$ 
bilinear, where we have an additional complication resulting 
from the fact that this bilinear necessarily gives 
rise to the production of charm quarks. The production 
of charm containing mesons from nucleon decay is of course kinematically 
forbidden. Thus our simplest nucleon decay process must involve an additional 
$c\rightarrow u$ conversion which will inhibit the decay width shown in 
Eq.(17), by an additional $G_F^2M_N^4$ factor;  
 where $G_F$ is the Fermi coupling constant.
By taking this additional complication into consideration it is found 
that the lower limit on $M_{\sigma}$ in the case of the 
$\langle\phi\rangle\langle\phi\rangle\sigma_{1.2}\sigma_{6.1}$
bilinear is:
\begin{eqnarray}
M_{\sigma}>\lambda^{3/4}\times 10^8{GeV}.
\end{eqnarray} 

We next consider the cubic terms, 
$b\sigma_i\sigma_j\sigma_j$ and
$\lambda\langle\phi\rangle\sigma_i\sigma_j\sigma_j$. 
The simplest nucleon decay diagrams arising from the former of these terms 
involves the decay of a nucleon into one meson, two leptons, and one 
antilepton. On the other hand the latter, Higgs containing, cubic term 
gives rise 
to decays involving the production
of two mesons, and one antilepton (see Fig.4), or in the special case of 
$\langle\phi\rangle\sigma_2\sigma_{1.2}\sigma_{1.2}$, one meson, 
two antileptons, and one lepton. 
The order of magnitude decay widths for the $b\sigma_i\sigma_j\sigma_j$ and
the $\lambda\langle\phi\rangle\sigma_i\sigma_j\sigma_j$ cubics are:
\begin{equation}
\Gamma \simeq {\cal O}\left(\frac{\lambda^6 b^2 M_N^{11}}
{M_{\sigma_j}^8 M_{\sigma_i}^4}\right)
\end{equation}
and
\begin{equation}
\Gamma \simeq {\cal O}\left(\frac{\lambda^8\langle\phi\rangle^2 M_N^{11}}
{M_{\sigma_j}^8 M_{\sigma_i}^4}\right),
\end{equation}
respectively. Note that we have included the dimensionless 
Yukawa coupling constants as a 
$\lambda^6\equiv \lambda_i^2\lambda_j^4$ factor. The nucleon 
decay processes resulting 
from these cubic terms are constrained by an experimental limit of around 
$\approx 10^{31}$yrs 
\cite{protondecay}. Therefore the lower limits on 
$M_{\sigma}$ for the $b\sigma_i\sigma_j\sigma_j$ and
the $\lambda\langle\phi\rangle\sigma_i\sigma_j\sigma_j$ cubics are:
\begin{equation}
M_{\sigma}>(\lambda^3 b/M_{\sigma})^{1/5}\times 10^6GeV
\end{equation}
and
\begin{equation}
M_{\sigma}>\lambda^{2/3}\times10^5{\normalsize GeV.}
\end{equation} 
respectively.

Finally we have the quartic terms
$\lambda\sigma_i\sigma_i^c\sigma_i\sigma_j$.
The simplest nucleon decay processes arising from these terms involves the 
decay of the nucleon into either, one meson, two
antileptons, and one lepton; or, depending on the scalars involved,
 two mesons and one antilepton. The decay width for these 
processes take the form
\begin{equation}
\Gamma\simeq {\cal O}\left(\frac{\lambda^{10}
M_N^{17}}{M_{\sigma_i}^{12}M_{\sigma_j}^4}\right),
\end{equation}
where $\lambda^{10}\equiv \lambda^2\lambda_i^6\lambda_j^2$.
These quartic nucleon decay processes are experimentally constrained 
by an approximately $10^{31}$
year lower limit, giving the constraint
\begin{equation}
M_{\sigma}>\lambda^{5/8}\times 10^4{GeV}.
\end{equation}

The simplest nucleon decay processes resulting from the quartic 
$\lambda\sigma_5\sigma_5\sigma_5\rho$, and the closely
related cubic $\lambda\sigma_5^a\sigma_5^b\sigma_5^b\langle\phi\rangle$,
entail the creation of a one
meson and three lepton product. The respective decay widths of these two
nucleon decay processes are shown below:
\begin{equation}
\Gamma\simeq {\cal O}\left(\frac{\lambda^{10}
M_N^{17}}{M_{\sigma_5}^{12}M_{\rho}^4}\right)
\end{equation}
\begin{equation} 
\Gamma\simeq {\cal O}\left(\frac{\lambda^{8}
\langle\phi\rangle^2 M_N^{11}}{M_{\sigma_5^a}^{4}M_{\sigma_5^b}^{8}}\right).
\end{equation}
Both of these decay widths have a dimensionless
$\lambda^8\equiv \lambda^2\lambda_5^6$ contribution, with the former
expression also containing a $\lambda^2$ contribution from the Yukawa constant associated
with the $\rho$ scalar.
For both potentials the strictest experimental constraint on 
$M_{\sigma_5,\rho}$ comes from bound 
neutron decay processes (see Fig.5), which have a lower lifetime limit of 
about $10^{31}$ years \cite{protondecay}. From this
constraint we obtain lower limits on $M_{\sigma, \rho}$ of 
\begin{equation}
M_{\sigma, \rho}>\lambda^{5/8}\times 10^4{GeV}
\end{equation}
and
\begin{equation}
M_{\sigma}>\lambda^{2/3}\times 10^5{GeV},
\end{equation}
respectively. This lower limit of $M_{\sigma, \rho}>\lambda^{5/8}
\times 10^4{GeV}$ is the lowest constraint on $M_{\sigma}$ for any of the 
$\Delta B=1$ scalar pairs listed in Eq.(12).

Before leaving these $\Delta B=1$ processes, we still have to consider the interesting case where one of the decay products is necessarily a right handed neutrino or a left handed antineutrino.
This occurs for all of the $\sigma_4$ containing potentials in Eq.(12). 
We assume the usual see-saw model where $\nu_R$ gains a large Majorana mass, 
M, which leads to the usual hierarchy of masses \cite{seesaw},
\begin{equation}
m_{\nu_R}\simeq M>>m>>m_{\nu_L}\simeq \frac{m^2}{M}
\end{equation}
where $m$ is the Dirac mass of the neutrinos, as given in the Dirac mass
term $m \bar{\nu}_L\nu_R$, and $M$ is the Majorana mass of the
neutrinos, as given in the Majorana mass term $M \bar{\nu}_R(\nu_R)^c$.  
The massiveness of this right handed neutrino means that any proton
decay producing such a particle will be highly suppressed, as the amount
of left hand right hand neutrino mixing will be very small. By assuming
 that the Dirac mass of the neutrino is around the same as the
Dirac masses of the other fermions, it is found that the suppression in
the decay width will be of order $\approx 10^{-14}$.

By considering this additional attenuation, 
our lower limits on $M_{\sigma}$ for the $\sigma_4$ containing interactions 
are found to reduce to; $\lambda\times 10^{12}$GeV for the
conjugate pair $\sigma_4$-$\sigma_{7.1}^c$; to 
 $(\lambda \mu/M_{\sigma}) \times 10^{12}$GeV for the quadratic
 $\sigma_4 \sigma_{7.1}$; to $\lambda^{3/4}\times 10^7$GeV for the 
quadratic $\sigma_{3.2} \sigma_4 \phi\phi$; to
 $\lambda^{5/8}\times 10^3$GeV for the
quartic $\sigma_{7.1}\sigma_{7.1}^c\sigma_{7.1}\sigma_4$; and to
 $\lambda^{5/8}\times 10^2$ GeV for the quartic
 $\sigma_4\sigma_4^c\sigma_4\sigma_{7.1}$.
For the scalar combination
$\sigma_4, \sigma_{7.1}$ our strongest constraint $M_{\sigma}>(\lambda\mu/M_{\sigma}) \times 10^{12}$GeV, has thus been reduced in comparison to the $(\lambda\mu/M_{\sigma}) \times 10^{16}$GeV constraint obtained for the quadratics
 in Eq.(12) without the $\sigma_4$ scalar.

The $\Delta B=2$ scalar potentials, which consist of cubic
and quartic terms are shown below:
\begin{eqnarray}
\sigma_1,\sigma_3&\rightarrow&\sigma_{1.1}\sigma_{3.1}\sigma_{1.1}\sigma_{3.1}\nonumber\\
&\rightarrow&\sigma_{1.1}\sigma_{3.2}\sigma_{1.1}\sigma_{3.2}\nonumber\\
&\rightarrow&\sigma_{1.1}\sigma_{3.3}\sigma_{1.1}\sigma_{3.3}\nonumber\\
&\rightarrow&\sigma_{1.1}\sigma_{3.4}\sigma_{1.1}\sigma_{3.4}\nonumber\\
&\rightarrow&\sigma_{1.2}\sigma_{3.1}\sigma_{1.2}\sigma_{3.1}\nonumber\\
&\rightarrow&\sigma_{1.2}\sigma_{3.2}\sigma_{1.2}\sigma_{3.2}\nonumber\\
&\rightarrow&\sigma_{1.2}\sigma_{3.3}\sigma_{1.2}\sigma_{3.3}\nonumber\\
&\rightarrow&\sigma_{1.2}\sigma_{3.4}\sigma_{1.2}\sigma_{3.4}\nonumber\\
\sigma_3,\sigma_7&\rightarrow&\sigma_{3.1}\sigma_{3.1}\sigma_{7.2}\nonumber\\
&\rightarrow&\sigma_{3.2}\sigma_{3.2}\sigma_{7.2}\nonumber\\
&\rightarrow&\sigma_{3.3}\sigma_{3.3}\sigma_{7.2}\nonumber\\
&\rightarrow&\sigma_{3.4}\sigma_{3.4}\sigma_{7.2}\nonumber\\
\sigma_4\sigma_7&\rightarrow&
\sigma_4\sigma_{7.1}\sigma_4\sigma_{7.1}\nonumber\\
&\rightarrow&\sigma_4\sigma_{7.2}\sigma_4\sigma_{7.2}\nonumber\\
\sigma_6,\sigma_7&\rightarrow&\sigma_{6.2}\sigma_{7.1}\sigma_{7.1}\nonumber\\
&\rightarrow&\sigma_{6.2}\sigma_{7.2}\sigma_{7.2}.
\end{eqnarray}
Unlike the $\Delta B=1$ processes, these $\Delta B=2$ processes will not
give rise to nucleon decay. In order to obtain constraints
on the scalar masses for these cases we must therefore compare the above 
interactions with $\Delta B=2$ experimental limits, such as binucleon decay 
measurements.
 
The cubic terms in Eq.(31), i.e. $b\sigma_i\sigma_j\sigma_j$, give rise to 
binucleon decay which in the simplest cases result in the production of two mesons. For example 
the neutron-neutron decay diagram resulting from the 
$\sigma_{3.1}\sigma_{3.1}\sigma_{7.2}$ process is shown in Fig.6. The decay
widths for these processes, using a dimensional approach, take the following 
form:
\begin{equation}
\Gamma\simeq{\cal O}\left(
 \frac{\lambda^6 b^2 M_N^{11}}{M_{\sigma_i}^4 M_{\sigma_j}^8}\right),
\end{equation}
where $M_N$ represents the mass of the
nucleons $\simeq 1$ GeV, and the dimensionless constant 
$\lambda^6\equiv \lambda_i^2\lambda_j^4$, where $\lambda_i$ and
$\lambda_j$ again represent the Yukawa couplings associated with
$\sigma_i$ and $\sigma_j$.
There is an approximate $10^{31}$ year \cite{protondecay} lower limit on these 
binucleon decay processes, thus for these trilinear terms we have a lower 
limit on $M_{\sigma}$ of
\begin{equation}
M_{\sigma}>(\lambda^3 b/M_{\sigma})^{1/5}\times 10^6{GeV.}
\end{equation}

The quartic terms in Eq.(31), i.e. $\lambda\sigma_i\sigma_j\sigma_i\sigma_j$, 
will give rise to binucleon decays which in the 
simplest cases will result 
in the creation of a meson and two antilepton product [see Fig.7]. The decay 
widths of these processes take the form
\begin{equation}
\Gamma\simeq{\cal O}\left(
 \frac{\lambda^{10} M_N^{17}}{M_{\sigma_i}^8 M_{\sigma_j}^8}\right),
\end{equation}
where $\lambda^{10}\equiv \lambda^2\lambda_i^4\lambda_j^4$. 
These decay processes are again constrained by 
an approximate $10^{31}$year limit, giving a lower limit on $M_{\sigma}$ of
\begin{equation}
M_{\sigma}>\lambda^{5/8}\times 10^4 GeV.
\end{equation}
For the $\sigma_4$ containing quartics we have an extra supression resulting 
from the production of two left handed antineutrinos. The lower limit on 
$M_{\sigma}$ for these quartics is thus reduced to:
\begin{equation}
M_{\sigma}>\lambda^{5/8}\times 10^2 GeV
\end{equation} 
This is the least stringent constraint on $M_{\sigma}$ which we have obtained. 
Therefore the $\sigma_4-\sigma_{7.2}$ scalar combination is 
the combination with the weakest constraint on $M_{\sigma}$; the 
$\sigma_4-\sigma_{7.1}$ combination is of course strictly constrained by its 
$\Delta B=1$ processes.

\begin{center}
{\bf 4. Conclusion:}
\end{center}

In this paper we have demonstrated how the observed charge quantisation
can be accounted for solely through classical constraints. In order to
obtain complete charge quantisation from classical constraints alone, we
extended the minimal standard model to include right handed neutrinos
and baryon number violation. We have effectively suggested that the necessity
of charge quantisation from classical constraints provides a strong
argument in favour of the existence of these baryon number violating processes.
We considered only simple extensions of the SM Yukawa interaction
where our new quark lepton interactions couple through new scalar particles
$\sigma$.

We know from experimental data on the decay of the proton and
decay of nuclei that baryon number violation is very much inhibited.
 Thus if these
new baryon number violating Yukawa interactions exist, the masses of the
associated 
scalars must be above a certain lower limit so as to evade detection by
present day experiments. Using a dimensional approach, these 
lower limits were calculated for all of our proposed scalars and scalar
combinations. 
From this exercise it was
found that the scalar pair with the lowest constraints on $M_{\sigma}$ is 
the $\sigma_4-\sigma_{7.2}$ combination which gives rise to the quartic term 
listed in Eq.(31).
This weakly constrained scalar combination is of interest 
as the possibility of these scalars appearing in low energy interactions 
is not ruled out.

As a result of the fact that present day 
theories on baryogenesis suggest that baryon number
violation must have occured in the early universe, it would be of some 
interest to investigate the implications of these baryon number violating 
processes on baryogenesis.

\newpage
\noindent
{\bf Figure 1:}  The proton decay diagram resulting from conjugate pair
$\sigma_{1.2}-\sigma_{3.2}^c$.
\vskip 5mm
\noindent
{\bf Figure 2:}  Proton decay resulting from the $\sigma_{1.1}\sigma_{3.1}$
quadratic.
\vskip 5mm
\noindent
{\bf Figure 3:}  Neutron decay resulting from the
$\langle\phi\rangle\sigma_2\sigma_{7.1}$ quadratic.
\vskip 5mm
\noindent
{\bf Figure 4:}  Proton decay resulting from the
$\langle\phi\rangle\sigma_{2.2}^c\sigma_{3.2}\sigma_{3.2}$ cubic.
\vskip 5mm
\noindent
{\bf Figure 5:}  Neutron decay resulting from the 
$\sigma_5\sigma_5\sigma_5\rho$ interaction.
\vskip 5mm
\noindent
{\bf Figure 6:}  Double Neutron decay resulting from the
$\sigma_{3.1}\sigma_{3.1}\sigma_{7.2}$ interaction.
\vskip 5mm
\noindent
{\bf Figure 7:}  Double proton decay resulting from the
$\sigma_{1.1}\sigma_{3.2}\sigma_{1.1}\sigma_{3.2}$ interaction.
\newpage
\vskip 5mm
\centerline{\epsfbox{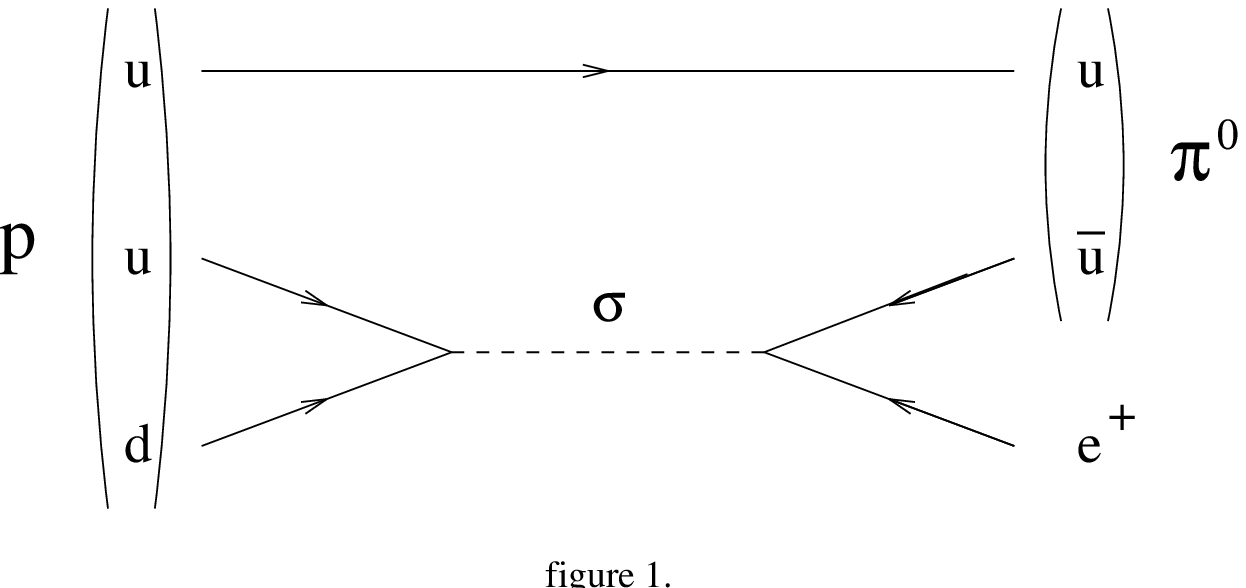}}
\vskip 50mm
\centerline{\epsfbox{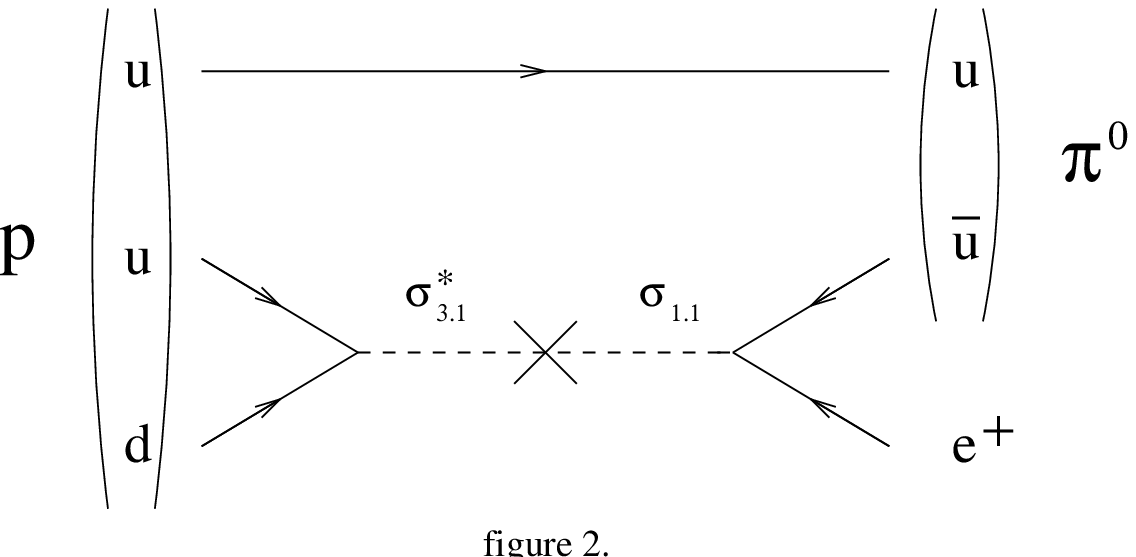}}
\newpage
\centerline{\epsfbox{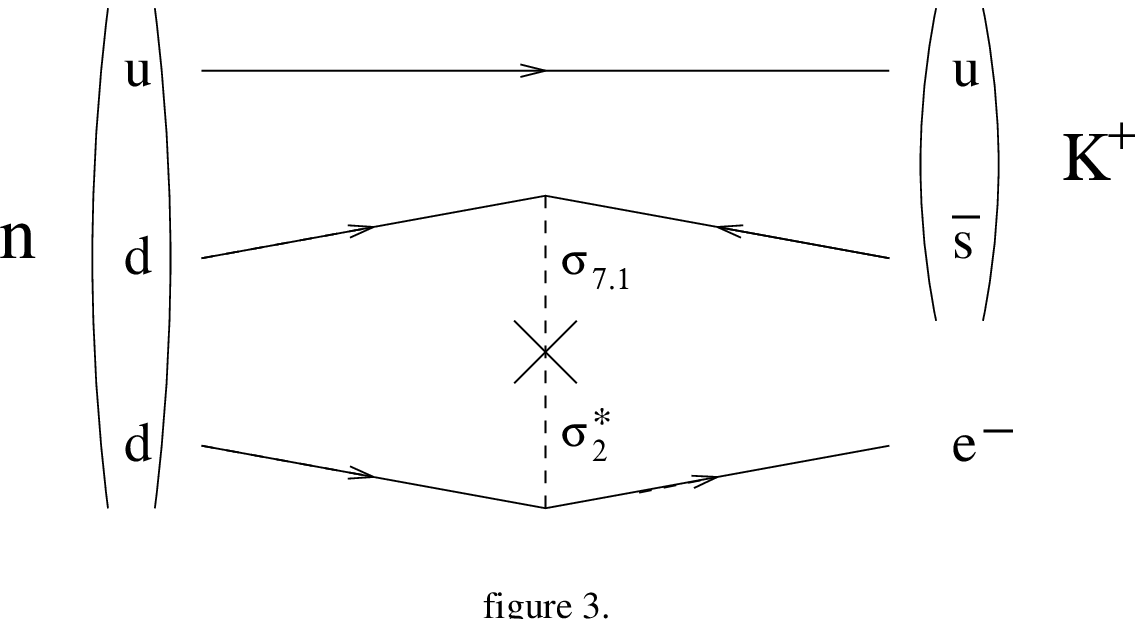}}
\vskip 50mm
\centerline{\epsfbox{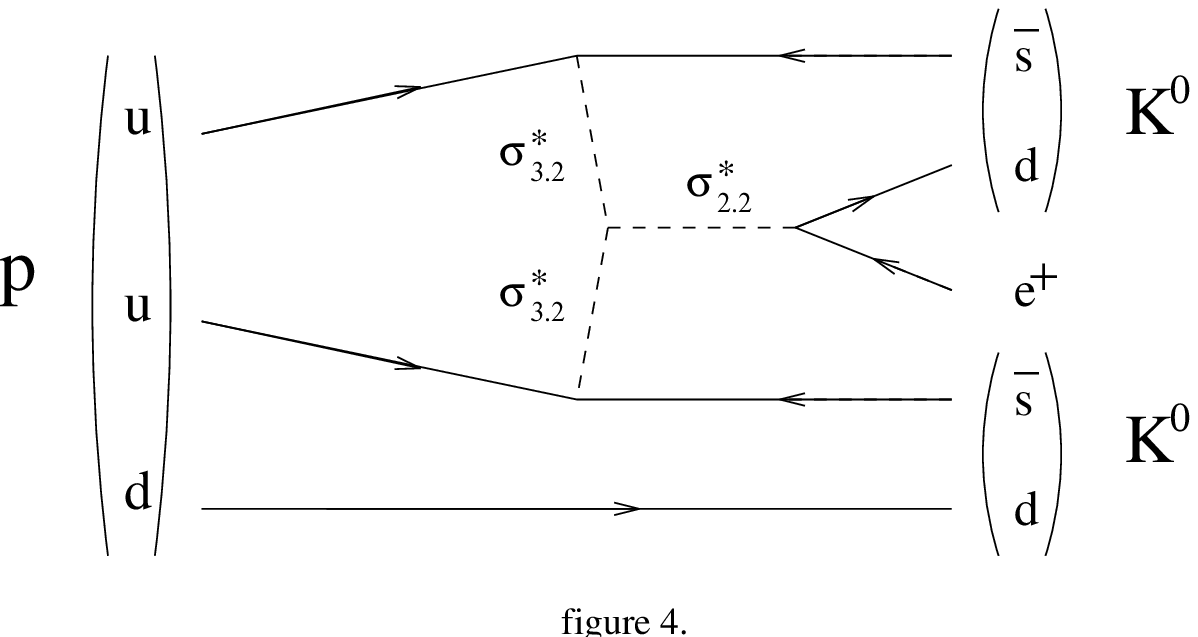}}
\newpage
\centerline{\epsfbox{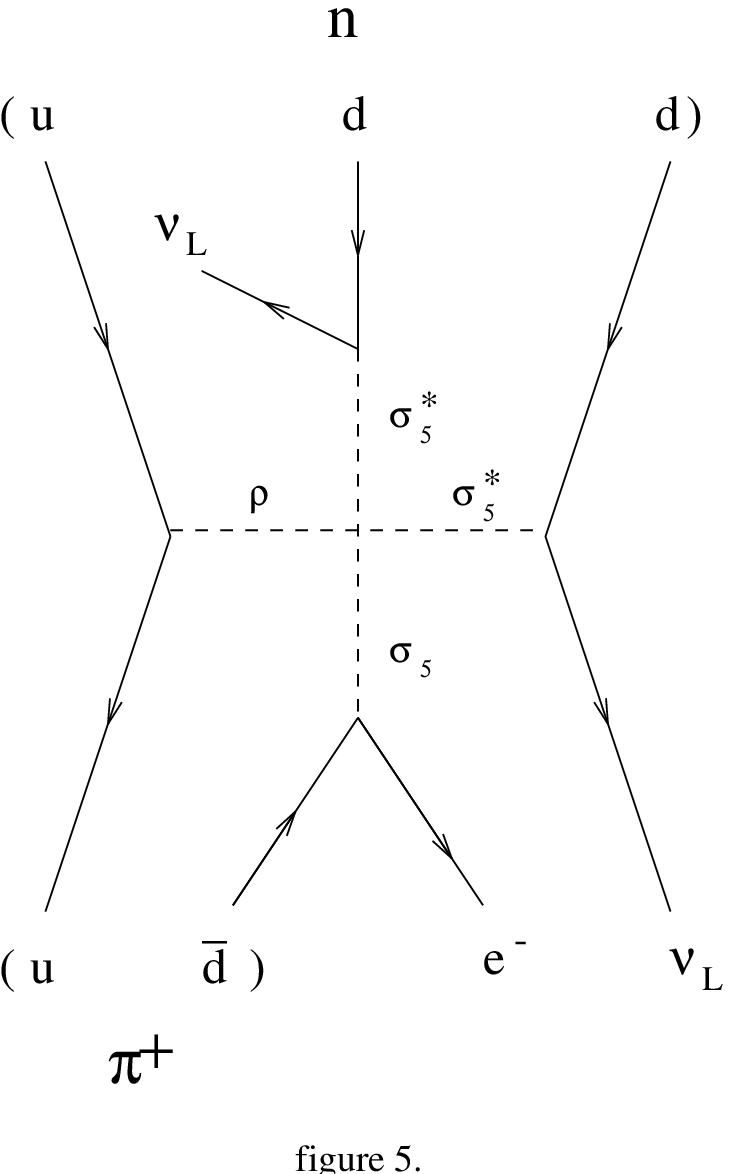}}
\newpage
\centerline{\epsfbox{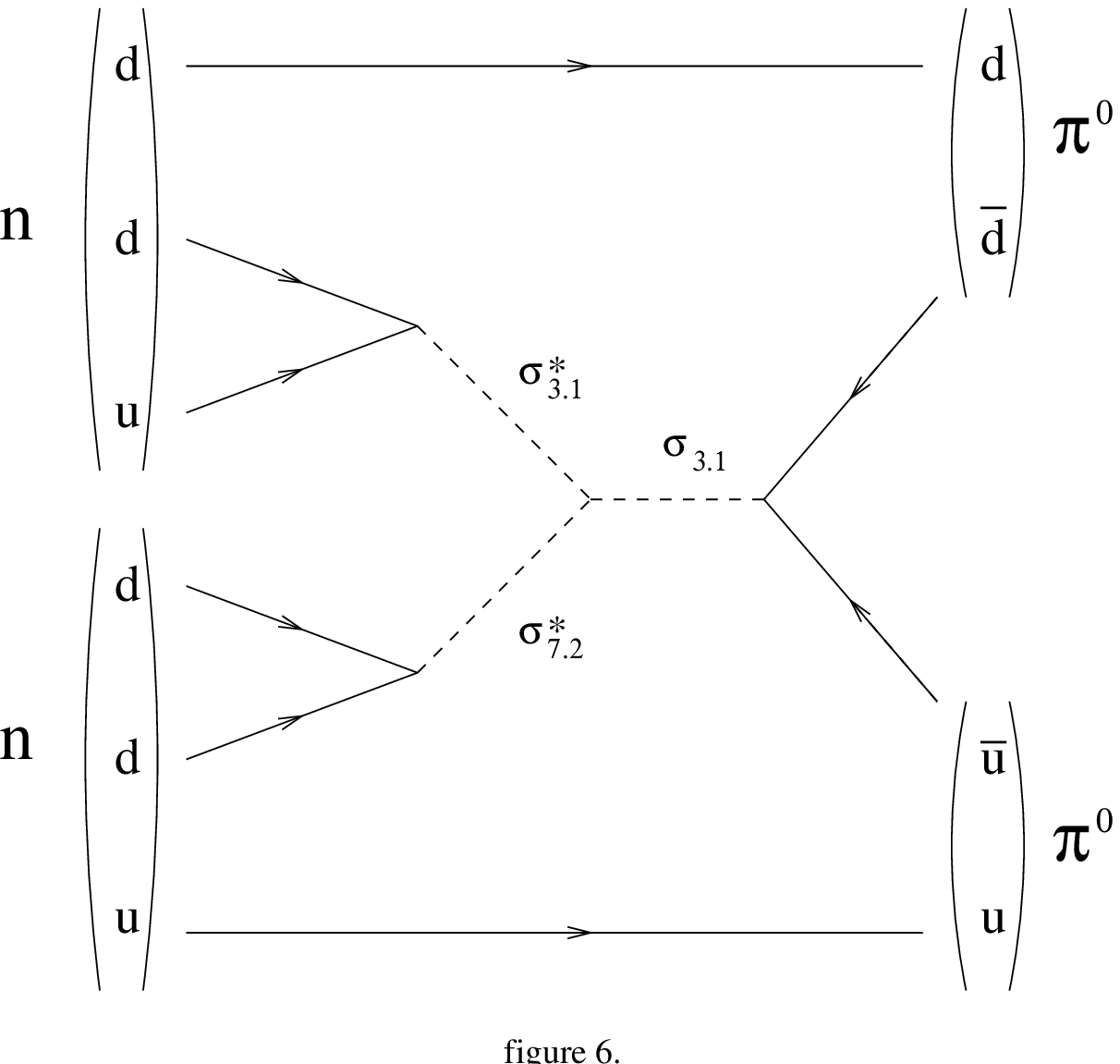}}
\vskip 50mm
\centerline{\epsfbox{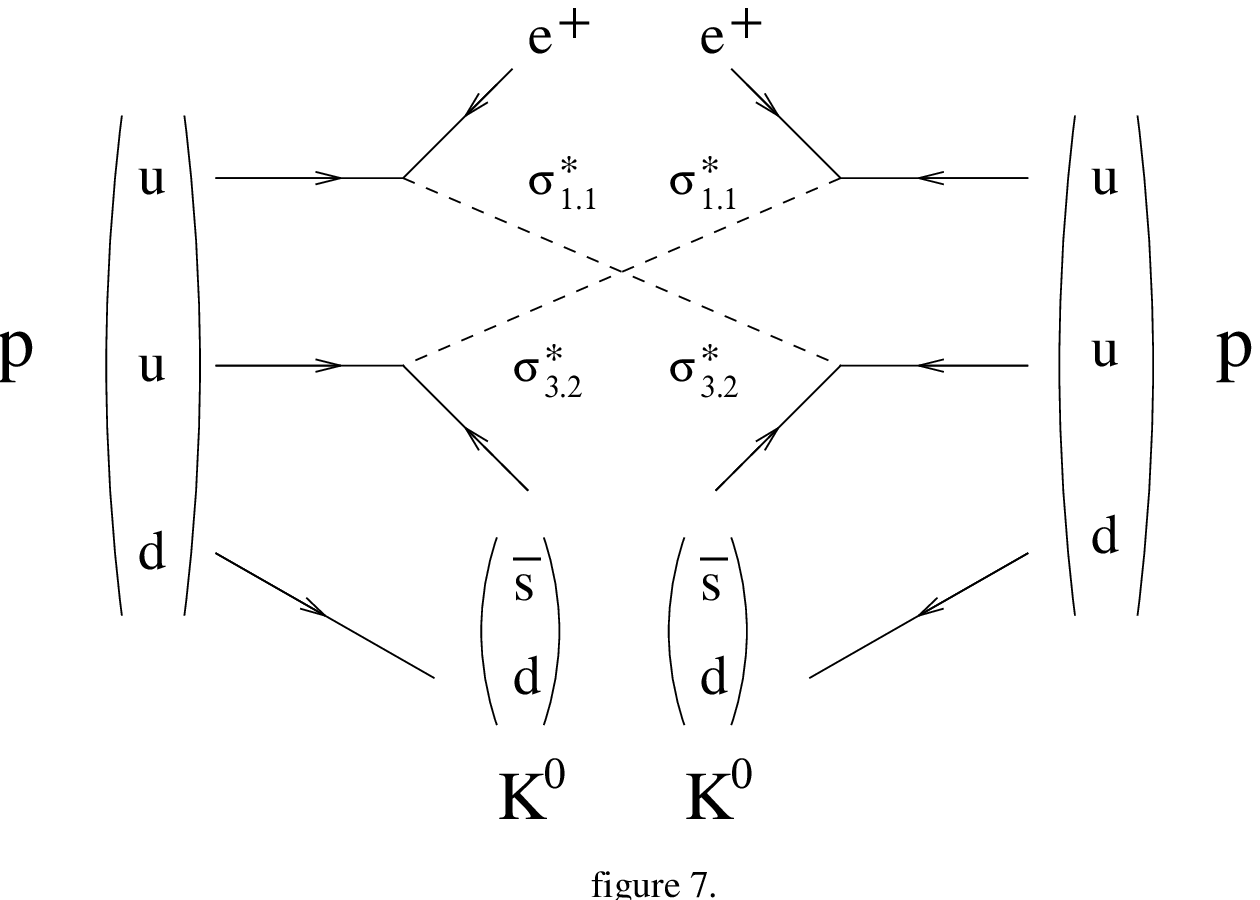}}
\end{document}